\documentclass[aps,pre,reprint,superscriptaddress,showpacs,floatfix]{revtex4-1}
\usepackage{amsmath}
\usepackage{amsfonts}
\usepackage{amssymb}
\usepackage{verbatim}
\usepackage{amsthm}
\usepackage{graphicx}
\usepackage{color}
\usepackage{booktabs}
\usepackage{array}

\newcommand{\figtext}[1]{\linespread{1.1} \small #1}

\newcommand{\fig}{Fig.\ }
\newcommand{\sect}{Sec.\ }
\newcommand{\eqn}{Eq.\ }
\newcommand{\eqns}{Eqs.\ }

\newcommand{\physrep}{Phys.\ Rep. }
\newcommand{\epl}{Europhys.\ Lett. }
\newcommand{\amjphys}{Am.\ J.\ Phys.}
\newcommand{\jphyscond}{J.\ Phys.: Condens. Matter}
\newcommand{\eurphysjb}{Eur.\ Phys.\ J.\ B}

\begin{document}


\title{Approach to equilibrium of diffusion in a logarithmic potential}
\author{Ori Hirschberg}
\author{David Mukamel}
\affiliation{Department of Physics of Complex Systems, Weizmann
Institute of Science, 76100 Rehovot, Israel}
\author{Gunter M. Sch{\"u}tz}
\affiliation{Theoretical Soft Matter and Biophysics, Institute of
Complex Systems, Forschungszentrum J\"ulich, 52425 J\"ulich,
Germany}


\date{\today}


\begin{abstract}
The late-time distribution function $P(x,t)$ of a particle diffusing
in a one-dimensional logarithmic potential is calculated for
arbitrary initial conditions. We find a scaling solution with three
surprising features: (i) the solution is given by two distinct
scaling forms, corresponding to a diffusive ($x\sim t^{1/2}$) and a
subdiffusive ($x \sim t^\gamma$ with a given $\gamma < 1/2$) length
scale, respectively, (ii) the overall scaling function is selected
by the initial condition, and (iii) depending on the tail of the
initial condition, the scaling exponent which characterizes the
scaling function is found to exhibit a transition from a
continuously varying to a fixed value.
\end{abstract}


 \maketitle

\section{Introduction}

There exist many physical systems whose temporal evolution is
described by a diffusion process in a one-dimensional logarithmic
potential. Examples include the denaturation process of DNA
molecules \cite{bar2009DNAMelting} and the temporal evolution of the
momentum distribution of cold atoms trapped in an optical lattice
\cite{CohenTannoudji,Zoller1996,LutzErgodicity2003}. Effective
logarithmic potentials also appear, for example, in models of
real-space condensation such as the zero-range process
\cite{evanszrpreview,GodrecheLuck2001ZetaUrn,ProbeParticles},
relaxation to equilibrium of long-range interacting gases
\cite{BouchetDauxoisRapid2005,CampaEtalLongRangeReview2009},
dynamics of the two-dimensional $XY$ model below the
Kosterlitz-Thouless transition \cite{BrayPersistence2000}, the ABBM
model for Barkhausen noise \cite{BarkhauseReview2006Lett}, and
dynamics of sleep-wake transitions during a night's sleep
\cite{LoEtAlSleepWake2002}.

The Fokker-Planck equation corresponding to diffusion in a
one-dimensional potential is
\begin{equation}\label{eq:FokkerPlanck}
\frac{\partial P(x,t)}{\partial t} = \frac{\partial}{\partial
x}\Bigl[V'(x)P(x,t) + \frac{\partial P(x,t)}{\partial x}\Bigr],
\end{equation}
where $P(x,t)$ is the probability distribution. We consider
potentials which increase logarithmically at large $x$,
\begin{equation}\label{eq:LogPotential}
V(|x|\gg 1) \sim b \log (|x|),
\end{equation}
and are regular at $x=0$. For simplicity, we have taken in these
equations the diffusion constant and the temperature to be equal to
1. Here we consider the case $b>1$, for which the system evolves
into a stationary state given by the normalizable Boltzmann
distribution
\begin{equation}\label{eq:EqDist}
P^*(x)  = \frac{1}{Z}e^{- V(x)} \sim \frac{1}{Z}x^{-b}
\end{equation}
where $Z = \int e^{-V(x)}dx$ is the normalization constant. For some
applications the variable $x$ is by definition non-negative, $x \geq
0$, as in the case of DNA denaturation where $x$ corresponds to the
length of a denaturated loop. In these cases, the equation has to be
supplemented by a boundary condition at the origin.

In this paper, we use a scaling analysis to study the long-time
evolution of the probability distribution towards the stationary
state. We find that the solution of \eqn (\ref{eq:FokkerPlanck})
relaxes to equilibrium via a universal scaling form which depends on
the potential only through its asymptotic form
(\ref{eq:LogPotential}). This scaling form exhibits several features
which are not typically found in scaling solutions
\cite{BarenblattBook}. (i) At large times, the equation exhibits two
distinct scaling regimes which we refer to as the large-$x$ and the
small-$x$ regimes. The two scaling functions yield, to leading order
in time, the distribution at \emph{any} point $x$. They join
smoothly at an intermediate scale $x_1(t)$, which grows with time.
(ii) The overall scaling solution (composed of both regimes) is not
unique. There exists a one parameter family of such solutions and
the appropriate solution is selected by the tail of the initial
distribution. The mechanism by which the initial condition selects
the eventual scaling solution is analogous to that encountered in,
e.g., fronts propagating into unstable states
\cite{BarenblattBook,VanSaarloosReview2003}. (iii) For a class of
initial conditions whose tails are sufficiently close to the
eventual steady-state distribution, the scaling solution is found to
be independent of the details of the initial condition. On the other
hand, the scaling function resulting from other initial conditions
varies continuously with the initial condition.

A large-$x$ scaling solution of \eqns
(\ref{eq:FokkerPlanck})--(\ref{eq:LogPotential}) has recently been
analyzed in
\cite{GodrecheLuck2001ZetaUrn,ProbeParticles,BarkaiKessler}, where
the dependence on the initial distribution has not been considered.
Although this analysis is valid for a rather broad class of initial
conditions, including compactly supported ones, other initial
conditions which may arise in various physical circumstances are
left out. The analysis presented here applies to all initial
conditions, and provides the scaling form in both the small-$x$ and
large-$x$ regimes.

The paper is organized as follows. In \sect \ref{sec:Scaling}, we
present our scaling analysis. The scaling solution for the case of
reflecting boundary conditions at the origin is considered in \sect
\ref{sec:ReflectingBC}, where a one-parameter family of solutions is
identified. In \sect \ref{sec:Selection}, we present a mechanism by
which a particular solution is selected by the initial condition,
and test it numerically. A generalization to other boundary
conditions is discussed in \sect \ref{sec:OtherBC}. Finally, in
\sect \ref{sec:conclusion}, we summarize our results.

\section{Scaling analysis\label{sec:Scaling}}
\subsection{Scaling solutions for reflecting boundary conditions\label{sec:ReflectingBC}}

We begin our analysis by introducing a function $G(x,t)$ defined as
\begin{equation}\label{eq:GDefinition}
P(x,t) = P^*(x)\bigl[1+G(x,t)\bigr].
\end{equation}
The deviation $P^*(x) G(x,t)$ from the steady state also satisfies
\eqn (\ref{eq:FokkerPlanck}), but its normalization is zero. For $x
\gg 1$, where the deviation of the potential from the logarithmic
form (\ref{eq:LogPotential}) is negligible, $G$ satisfies
\begin{equation}\label{eq:FokkerPlanckForGLargeX}
\frac{\partial G(x,t)}{\partial t} = -\frac{b}{x}\frac{\partial
G(x,t)}{\partial x} + \frac{\partial^2G(x,t)}{\partial x^2}.
\end{equation}
To be specific, we consider the case where the equation is defined
on the positive real axis, $x\geq 0$, with a reflecting boundary
condition at $x=0$. This implies $\partial G(x=0,t)/\partial x = 0$.
We later comment on other boundary conditions.

The scaling solution of \eqn (\ref{eq:FokkerPlanckForGLargeX}) may
in fact be obtained by an exact solution of \eqns
(\ref{eq:FokkerPlanck})--(\ref{eq:LogPotential}). This is done by
transforming the Fokker-Planck equation into an imaginary-time
Schr{\"o}dinger equation via the transformation $P(x,t) =
e^{-V(x)/2}\psi(x,t)$ \cite{Risken}. The resulting equation for the
``wavefunction'' $\psi$ is
\begin{equation}\label{eq:QMSchrodinger}
\frac{\partial \psi(x,t)}{\partial t} =
\frac{\partial^2\psi(x,t)}{\partial x^2} - V_s(x) \psi(x,t)
\end{equation}
with the Schr{\"o}dinger potential $V_s(x) = \frac{V'(x)^2}{4} -
\frac{V''(x)}{2}$. For a potential of the form
(\ref{eq:LogPotential}) this gives
\begin{equation}\label{eq:ExactSchrodingerPotential}
V_s(x\gg 1) \sim \gamma/x^2
\end{equation}
with $\gamma = \frac{b}{2}(\frac{b}{2}+1)$. For large $x$, this
equation describes the well studied problem of a quantum particle
moving in a repulsive inverse square potential
\cite{QMInverseSquareAmJPhys2006}. The solution of this equation may
be found by expanding it in eigenfunctions, $\psi(x,t) = \int dk \,
a_k e^{-k^2 t}\psi_k(x)$, where $\psi_k(x)$ can be expressed in
terms of Bessel functions. The long-time scaling behavior is then
obtained by studying the small-$k$ behavior of the amplitudes $a_k$.
Carrying out this expansion, we find that any localized initial
condition for the function $G$ evolves at long times to
\begin{equation}\label{eq:ScalingFunctionBeta1}
G(x,t) \sim t^{-1}f_1\Bigl(\frac{x}{\sqrt{t}}\Bigr)\quad \text{ with
}\quad f_1(u) = u^{b+1}e^{-u^2/4}.
\end{equation}
The analysis is rather lengthy and will be presented elsewhere
\cite{LongVersion}. Below we derive the long-time solution directly
by assuming a scaling form. We make use of the result
(\ref{eq:ScalingFunctionBeta1}) only in relating the appropriate
scaling solution to the initial condition.

Let us now present the scaling solution of this equation and outline
its derivation. As we demonstrate below, two length scales emerge at
large times: a large-$x$ regime with $x \sim t^{1/2}$, and a
small-$x$ regime, $x\sim t^\gamma$, with a $b$-dependent $\gamma$
satisfying $\gamma < 1/2$. Starting with the small-$x$ regime, we
consider a scaling solution of the form
\begin{equation}\label{eq:SmallXScalingAnsatz}
G(x,t) = t^{-\delta}g(z),\quad z = \frac{x}{t^{\gamma}}
\end{equation}
with some function $g$ and exponents $\gamma$ and $\delta$.
Substituting (\ref{eq:SmallXScalingAnsatz}) in \eqn
(\ref{eq:FokkerPlanckForGLargeX}) yields
\begin{equation}\label{eq:SmallXScalingODE}
g''(z) - \frac{b}{z}g'(z) = -[\gamma z g'(z) + \delta
g(z)]t^{-(1-2\gamma)}.
\end{equation}
The right-hand side of this equation may be neglected as long as
$\gamma < 1/2$, yielding the solution
\begin{equation}\label{eq:smallXScalingSolution}
g(z) = \tilde{C} + C z^{b+1},
\end{equation}
where $C$ and $\tilde{C}$ are integration constants. Thus, for $z\ll
1$ one has $G(zt^\gamma,t) \sim t^{-\delta} \tilde{C}$. In fact,
since for $z \ll 1$ this solution satisfies the boundary condition
at $x=0$, it is valid down to $x=0$.

In order to determine $\gamma$ and $\delta$ one needs the solution
in both small-$x$ and large-$x$ regimes. We thus consider a
different scaling function for $x \sim t^{1/2}$,
\begin{equation}\label{eq:ScalingAnsatz}
G(x,t) = t^{-\beta} {f}(u),\quad u = \frac{x}{t^{1/2}}
\end{equation}
where the scaling exponent $\beta$ and the function $f(u)$ are to be
determined. Substituting (\ref{eq:ScalingAnsatz}) in \eqn
(\ref{eq:FokkerPlanckForGLargeX}) yields a family of ordinary
differential equations for $f(u)$, parameterized by $\beta$:
\begin{equation}\label{eq:ScalingODE}
f''+\Bigl(\frac{u}{2}-\frac{b}{u}\Bigr)f' + \beta f = 0.
\end{equation}
Requiring that the small-$x$ and large-$x$ solutions join smoothly
at an intermediate scale
\begin{equation}\label{eq:FrontBoundaryRange}
t^\gamma \ll x_1(t) \ll t^{1/2},
\end{equation}
one concludes from (\ref{eq:SmallXScalingAnsatz}),
(\ref{eq:smallXScalingSolution}) and (\ref{eq:ScalingAnsatz}) that
\begin{equation}\label{eq:ODEBoundaryCondition}
f(u \ll 1) \sim C u^{b+1},
\end{equation}
and
\begin{equation}\label{eq:SmallXDelta}
\delta = \beta + (b+1)({\textstyle \frac{1}{2}}-\gamma).
\end{equation}
The solution of \eqn (\ref{eq:ScalingODE}) which satisfies
(\ref{eq:ODEBoundaryCondition}) is \cite{AbramowitzStegun}
\begin{equation}\label{eq:ODESolution2}
f(u) = C
u^{b+1}\,_1\!F_1\left(\frac{1+b+2\beta}{2};\frac{b+3}{2};-\frac{u^2}{4}\right)
,
\end{equation}
where $_1\!F_1$ is the hypergeometric function. For small and large
arguments $f(u)$ satisfies \cite{AbramowitzStegun}
\begin{equation}\label{eq:AsymptoticF}
f(u) \sim \left\{
\begin{array}{ll} u^{b+1}\phantom{e^{-\frac{u^2}{4}}}
& \text{ for } u \ll 1 \\
Bu^{-2\beta}\phantom{e^{-\frac{u^2}{4}}} & \text{ for } u \gg 1,
\beta \neq 1,2,3,\ldots \\
Bu^{b+2\beta-1}e^{-\frac{u^2}{4}} & \text{ for } u \gg 1, \beta =
1,2,3,\ldots
\end{array} \right.,
\end{equation}
where $B$ is a known constant which depends on $b$ and $\beta$.

Once the small-$x$ and large-$x$ scaling functions are known,
conservation of probability enables us to determine the scaling
exponent $\gamma$ and the integration constant $\tilde{C}$. We
evaluate the normalization condition $\int_0^\infty P^*(x)G(x,t) dx
= 0$ by splitting the integral into two domains, $0 \leq x \leq
x_1(t)$ and $x_1(t)\leq x \leq \infty$. Using the solutions found in
the two domains and evaluating the integrals to leading order in $t$
we obtain
\begin{equation}\label{eq:SmallXGamma}
\gamma = \frac{1}{b+1} \quad \text{ and } \quad \delta = \beta +
\frac{b-1}{2}.
\end{equation}
Also,
\begin{equation}\label{eq:SmallXConstant1}
\tilde{C} = -\frac{C}{Z}\int_0^\infty u^{-b}f(u)du =
-\frac{C}{Z}\cdot\frac{2(b+1)}{2\beta+b-1}.
\end{equation}

\begin{figure}
  \center
  \includegraphics[width = 0.4\textwidth]{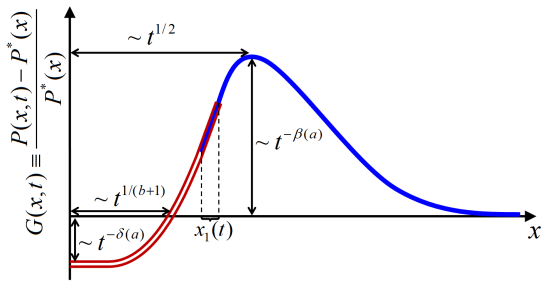}
  \caption{ \label{fig:SchematicScaling}\figtext{(Color online) A schematic
  representation of the scaling solution $G(x,t)$ (see \eqn (\ref{eq:FinalScalingAnsatz}))
  at a given late time $t \gg 1$ (not drawn to scale).
  The red double line represents the
  small-$x$ scaling form $g(x/t^{1/(b+1)})t^{-\delta}$ while the blue
  solid line represents the large-$x$ scaling form
  $f(x/t^{1/2})t^{-\beta}$. The interval on which the two solutions
  overlap (\eqn (\ref{eq:FrontBoundaryRange})) is labeled $x_1(t)$.
  }}
\end{figure}

Summarizing the results of the scaling analysis, we find that the
solution of \eqns (\ref{eq:FokkerPlanck})--(\ref{eq:LogPotential})
is, to leading order in $t$,
\begin{equation}\label{eq:FinalScalingAnsatz}
P(x,t) \approx P^*(x) +  P^*(x) \cdot \left\{
\begin{array}{ll} g\bigl(\frac{x}{t^{\gamma}}\bigr)
t^{-\delta},  & x\leq x_1(t) \\ \\
f \bigl(\frac{x}{t^{1/2}}\bigr)t^{-\beta}, & x\geq x_1(t),
\end{array} \right.,
\end{equation}
where $f$ and $g$ are given in \eqns
(\ref{eq:smallXScalingSolution}) and (\ref{eq:ODESolution2}),
$\gamma$ and $\delta$ are given in (\ref{eq:SmallXGamma}), and
$x_1(t)$ satisfies (\ref{eq:FrontBoundaryRange}). We thus obtain a
one-parameter family of scaling solutions parameterized by $\beta$.
Below we denote a member of this family by $P_\beta(x,t) \equiv
P^*(x)[1 + G_\beta(x,t)]$. The overall form of these scaling
solutions for $G$ is given schematically in \fig
\ref{fig:SchematicScaling}. For small-$x$, the function is flat,
with a value that approaches zero as $t^{-\delta}$. For $x \geq
x_1(t)$, it exhibits a peak at $x\sim\sqrt{t}$, whose height scales
as $t^{-\beta}$. Thus, as time progresses, the peak shrinks and
moves to the right, and $G$ approaches its steady-state value
$G(x,t\to\infty) = 0$.

\subsection{Selection mechanism\label{sec:Selection}}

To conclude our analysis, we argue now that the parameter $\beta$ is
selected by the tail of the initial distribution. To this end, we
consider initial distributions of the form
\begin{equation}\label{eq:InitialAsymptotics}
G_0(x \gg 1) \sim A x^{-a}
\end{equation}
where $a>b-1$ and $A$ are parameters. Initial conditions $G_0(x)$
which decay faster than algebraically correspond to $a=\infty$.
Note, however, that localized initial \emph{distributions} $P(x,0)$
correspond to $a=0$. Since the dynamical process of \eqn
(\ref{eq:FokkerPlanck}) is diffusive, one would naively expect the
tail of the distribution to remain unchanged for $x\gg \sqrt{t}$.
Using the asymptotics (\ref{eq:AsymptoticF}), this suggests that
$\beta = a/2$. The parameter $C$ is then given by $C=A/B$.
Substituting the value of $B$ yields \cite{AbramowitzStegun}
\begin{equation}\label{eq:FinalConstant}
C = \frac{\Gamma(1-a/2)}{2^{1+b+a}\Gamma(\frac{3+b}{2})}\, A.
\end{equation}

This naive argument is found to be valid only for $a<2$. For $a>2$
we make use of \eqn (\ref{eq:ScalingFunctionBeta1}) to demonstrate
that the correct scaling solution is given by $\beta = 1$. This is
done by analyzing the stability of a scaling solution with a given
$\beta$. We thus consider a perturbation $\delta P(x,t)$ around the
scaling solution (\ref{eq:FinalScalingAnsatz}):
\begin{equation}\label{eq:LocalDisturbance}
P(x,t) - P^*(x) = P^*(x) G_\beta(x,t) + \delta P(x,t),
\end{equation}
which is initially \emph{localized} in $x$, such as a function with
compact support. Normalization dictates that $\int \delta P dx = 0$.
This perturbation satisfies \eqns
(\ref{eq:FokkerPlanck})--(\ref{eq:LogPotential}).

\begin{figure}
  \center
  \includegraphics[width = 0.48\textwidth]{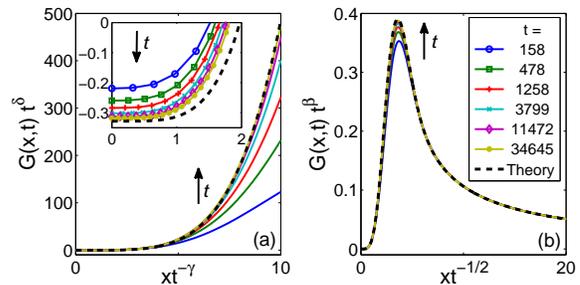}
  \caption{\label{fig:NumericalScaling}\figtext{(Color online) The (a) small-$x$ and (b)
  large-$x$ scaling collapse obtained by numerically integrating
  \eqn (\ref{eq:ZRPMasterEq}) for $b=3.5$ and $a=1$.
  Different curves correspond to
  different times in increasing order in the arrow direction.
  The dashed curve is the theoretical function
  (\ref{eq:FinalScalingAnsatz}), with no fitting parameter. Inset shows
  a magnified region around the origin.
  }}
\end{figure}

The scaling solution $G_\beta$ is stable to such perturbations and
will dominate the approach to equilibrium only if at late times
$\delta P(x,t)$ is negligible compared to it. The exact solution of
\eqn (\ref{eq:FokkerPlanck}) reveals that the scaling solution
(\ref{eq:ScalingFunctionBeta1}) to which a localized initial
condition converges at long times is of the form
(\ref{eq:FinalScalingAnsatz}) with $\beta = 1$ \cite{LongVersion}.
This can be understood heuristically by noting that $\beta = 1$ is
the most localized of all scaling solutions, see
(\ref{eq:AsymptoticF}). Therefore, the scaling form at large-$x$ of
\eqn (\ref{eq:LocalDisturbance}) is given by
\begin{multline}
P(x,t) - P^*(x) = P^*(x) G_\beta(x,t) + \delta
P(x,t)   \\
\approx P^*(x) \Bigl[
t^{-\beta}f_\beta\Bigl(\frac{x}{\sqrt{t}}\Bigr) +
t^{-1}f_1\Bigl(\frac{x}{\sqrt{t}}\Bigr)\Bigr],
\end{multline}
where $f_\beta$ is the solution (\ref{eq:ODESolution2})
corresponding to $\beta$. For $\beta < 1$ the second term on the
right-hand side is negligible compared to the first, and therefore
the $f_\beta$ solution is stable.
On the other hand, for $\beta
> 1$ the second term is dominant and the scaling solution
is given by $f_1$. Thus,
\begin{equation}\label{eq:FinalBeta}
\beta = \beta(a) = \left\{
\begin{array}{ll} \frac{a}{2} & \text{if } a<2 \\
1 & \text{if } a>2
\end{array} \right..
\end{equation}
The exact solution of the Fokker-Planck equation further reveals
that when $a=2$ there are logarithmic corrections to \eqn
(\ref{eq:FinalScalingAnsatz}) \cite{LongVersion}. Note that for
$a>2$, the constant $C$ is not given by (\ref{eq:FinalConstant}) and
it depends on the details of the initial condition. The large-$x$
scaling solution for localized initial distributions (corresponding
to $a=0$) agrees with the results previously obtained in
\cite{GodrecheLuck2001ZetaUrn,ProbeParticles,BarkaiKessler}.
Knowledge of the scaling solution for other initial conditions is
often required for calculating correlation functions of physical
interest. Applications of this approach to physical examples where
the late time behavior is determined by the initial conditions will
be presented elsewhere \cite{LongVersion}.

\begin{figure}
  \center
  \includegraphics[width = 0.3\textwidth]{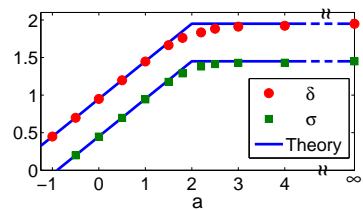}
  \caption{\label{fig:DeltaAndSigma}\figtext{The scaling exponents
  $\delta$ and $\sigma$ obtained by numerical integration of
  (\ref{eq:ZRPMasterEq}), compared with theoretically predicted
  values. Here $b=2.9$ and $A=1$. The transition at $a=2$ is readily
  seen.
  }}
\end{figure}

In order to check the applicability of the scaling solution found in
this analysis, we studied numerically the evolution of a single-site
zero-range process at criticality \cite{evanszrpreview}. In this
process, particles hop into a site with a constant rate 1, and hop
out of this site with rate $w(n) = 1+b/n$, where $n$ is the number
of particles in the site. The occupation number probability
distribution $P(n,t)$ satisfies the master equation
\begin{equation}\label{eq:ZRPMasterEq}
\frac{\partial}{\partial t}P(n) = P(n\!-\!1)  + w(n\!+\!1)P(n\!+\!1)
- [1+w(n)]P(n).
\end{equation}
This is a discrete version of
(\ref{eq:FokkerPlanck})--(\ref{eq:LogPotential}). Its steady state
is given by $P^*(n) \propto [w(1)\ldots w(n)]^{-1} \sim n^{-b}$. We
have studied the relaxation to the steady state starting from
various initial conditions. In \fig \ref{fig:NumericalScaling} we
display the results obtained for an initial condition with $a=1$. A
very good agreement with the predicted scaling functions is obtained
at long times. To investigate the dependence (\ref{eq:FinalBeta}) of
the scaling exponents on the initial condition, we have measured
$G(0,t)$ which is predicted to decay to zero as $t^{-\delta}$
(\ref{eq:SmallXGamma}). As an independent measure, we calculated
$\langle n(t) \rangle$, which is predicted to decay to its
equilibrium value $\langle n \rangle_{\text{eq}}$ as $t^{-\sigma}$
with $\sigma = \beta+b/2-1$. This can be easily obtained using the
scaling function (\ref{eq:FinalScalingAnsatz}):
\begin{equation}
\langle n(t) \rangle-\langle n \rangle_{\text{eq}} \approx
t^{-(\beta+b/2-1)}C\int_0^\infty u^{1-b}f(u)du.
\end{equation}
In \fig \ref{fig:DeltaAndSigma} we compare numerical measurements of
$\delta$ and $\sigma$ to the theoretical predictions and find a very
good agreement, both for $a<2$ and $a>2$.

\subsection{Other boundary conditions\label{sec:OtherBC}}
The analysis presented above may be extended to consider other
boundary conditions as well. For example, for an absorbing boundary
condition at $x=0$, namely $P(0,t) = 0$, the stationary distribution
vanishes. Defining $G(x,t)$ by
\begin{equation}\label{eq:GDefinitionAbsorbing}
P(x,t) = e^{-V(x)}G(x,t)
\end{equation}
and repeating the derivation outlined above for the evolution of
$G$, we find that in the scaling limit it takes the same form
(\ref{eq:ScalingAnsatz}) as before, with $f$ unchanged, $\gamma = 0$
and a different function $g(z)$ which can be calculated explicitly
\cite{LongVersion}. Equation (\ref{eq:FinalBeta}), which yields the
relation between $\beta$ and $a$, remains valid. However, since $G$
is defined differently in this case, the same initial distribution
$P(x,0)$ would correspond to different values of $a$ depending on
whether the boundary condition is reflecting or absorbing. Other
boundary conditions, and the case where the equation is defined on
the entire $x$-axis, may be treated similarly \cite{LongVersion}.

\section{Conclusion\label{sec:conclusion}}

In summary, the approach to equilibrium of a diffusion process in a
logarithmic potential is analyzed in the scaling limit and is shown
to exhibit uncommon and interesting features. These include the
existence of two characteristic length scales, the fact that the
scaling function depends on the tail of the initial distribution,
and the non-analytic way the scaling exponents varies with the
initial condition. The mechanism by which the scaling exponents are
selected is similar to the one encountered in problems of velocity
selection in propagating fronts
\cite{BarenblattBook,VanSaarloosReview2003}. Here, however, unlike
problems of front propagation, the evolution equation is linear,
although inhomogeneous. This facilitates the explicit demonstration
of the selection mechanism.

\begin{acknowledgments}
We thank A.\ Amir, A.\ Bar, O.\ Cohen, J.-P.\ Eckmann, and M.\ R.\
Evans, for useful discussions and comments on the manuscript. This
work was supported by the Israel Science Foundation (ISF).
\end{acknowledgments}

\bibliographystyle{apsrev4-1}
\bibliography{log_diff_long}

\end{document}